\documentclass[aps,preprint,amsmath,amssymb]{revtex4-1}
\usepackage{slashed}
\usepackage{graphicx}
\usepackage{color}
\newcommand{\nn}{\nonumber}
\newcommand{\bd}{\begin{document}}
\newcommand{\ed}{\end{document}}
\newcommand{\bc}{\begin{center}}
\newcommand{\ec}{\end{center}}
\newcommand{\be}{\begin{eqnarray}}
\newcommand{\ee}{\end{eqnarray}}
\newcommand{\ba}{\begin{array}}
\newcommand{\ea}{\ed{array}}
\newcommand{\strich}[1]{#1  \! \! \slash}
\newcommand{\eqn}{\global\def\theequation}
\newcommand{\sw}{sin^2 \theta_W}
\newcommand{\fbd}{f_B}
\renewcommand{\thefootnote}{\alph{footnote}}
\newcommand{\se}{\section}
\newcommand{\sse}{\subsection}
\newcommand{\bi}{\bibitem}
\def\figcap{\section*{Figure Captions\markboth
     {FIGURECAPTIONS}{FIGURECAPTIONS}}\list
     {Figure \arabic{enumi}:\hfill}{\settowidth\labelwidth{Figure 999:}
     \leftmargin\labelwidth
     \advance\leftmargin\labelsep\usecounter{enumi}}}
\let\endfigcap\endlist \relax
\def\reflist{\section*{References\markboth
     {REFLIST}{REFLIST}}\list
     {[\arabic{enumi}]\hfill}{\settowidth\labelwidth{[999]}
     \leftmargin\labelwidth
     \advance\leftmargin\labelsep\usecounter{enumi}}}
\let\endreflist\endlist \relax

\def\Journal#1#2#3#4{{#1} {{\bf #2},} {#4} {(#3)}}
\def\NCA{Nuovo Cimento}
\def\NIM{Nucl. Instrum. Methods}
\def\NIMA{{Nucl. Instrum. Methods} A}
\def\NP{{Nucl. Phys.} }
\def\NPB{{Nucl. Phys.} B }
\def\NPA{{Nucl. Phys. A}}
\def\PLB{{Phys. Lett.}  B}
\def\PL{{Phys. Lett.}}
\def\PPSA{{Proc. Phys. Soc.} A}
\def\PRP{{ Phys. Rep.}}
\def\PRL{ Phys. Rev. Lett.}
\def\PR{{Phys. Rev.}}
\def\PRD{{Phys. Rev.} D}
\def\PRC{{Phys. Rev.} C}
\def\ZP{{Z. Phys.}}
\def\ZPC{{Z. Phys. C}}
\def\EPJ{{Eur. Phys. J.}}
\def\EPJC{{Eur. Phys. J.} C}
\def\ZPA{{Z. Phys.} A}
\def\MPL{{Mod. Phys. Lett.}}
\def\MPLA{{Mod. Phys. Lett.} A}
\def\CPC{Comput. Phys. Commun.}
\def\JHEP{{J. High Energy Phys.}}
\def\JPG{{J. Phys. G.}}
\def\SJNP{Sov. J. Nucl. Phys.}
\def\NCA{ Nuovo Cimento}
\def\NIM{ Nucl. Instrum. Methods}
\def\NIMA{{ Nucl. Instrum. Methods} A}
\def\NP{{ Nucl. Phys.}}
\def\ANP{{Adv. Nucl. Phys.}}
\def\CPC{{Comput. Phys. Commun.}}

\begin{document}
\title{Time-like electromagnetic form factors of 
$\Lambda,~\Sigma$ and $\Xi^{+}$ in a light-front quark model}

\author{Chong-Chung Lih\footnote{cclih123@gmail.com} and 
Chao-Qiang Geng\footnote{cqgeng@ucas.ac.cn}}
\affiliation{
Chongqing University of Posts and Telecommunications, 
Chongqing, 400065, China\\
School of Fundamental Physics and Mathematical Sciences,
Hangzhou Institute for Advanced Study, UCAS, Hangzhou 310024, China}

\date{\today}

\begin{abstract}
We use the light front quark model to investigate the form factors 
in the $e^{+}e^{-} \to B\bar{B}$ collision proceses with $B=\Lambda,~\Sigma$ and $\Xi$. 
These form factor behaviors are calculated  
based on the Bethe-Salpeter formalism with $q^{+} > 0$ to effectively account  
for non-valence contributions. 
We show that our results of the $q^2$-dependent form factors closely match the BESIII data. 
In particular, we obtain that $|G_{eff}| = (0.921,~0.098,~0.189$) and 
$R =|$$G_E\over G_M$$|= (0.97,~0.89$,~$0.936$) for 
$e^{+} e^{-}\to \Lambda \bar{\Lambda}$,~$\Sigma^{+} \Sigma^{-},~\Xi^{+} \Xi^{+})$ 
with $q^2 = (5.74,~6.0,~7.0)$ GeV$^2$.

\end{abstract}

\maketitle

\se{Introduction}

Understanding the internal structure of hadrons has always been a great challenge. 
Quantum chromodynamics (QCD) uses quarks and gluons (i.e., QCD degrees of freedom) 
to describe the structure of hadrons and their interactions. 
We are accustomed to believe that the motion of quarks and gluons inside hadrons 
will change under the action of strong fields in nuclear matter, 
and such changes are expected to be reflected in the electromagnetic 
and weak structures of nucleons and other baryons \cite{qcd1,qcd2,qcd3}. 
A commonly used parametrization for the electromagnetic structure of 
hadrons is the electromagnetic form factor (EMFF), 
which is an observable quantity of non-perturbative QCD 
and is also a key to understand bound-state QCD effects. 
Therefore, the EMFF can be used to explore the internal structure 
of baryons and experimentally detected through the interaction 
of hadrons with virtual photons.

Another possibility for revealing the electromagnetic structure of baryons 
is the $e^{+}e^{-}$ scattering. It allows us to enter the time-like regime 
($q^2 = -Q^2 \geq 0$), which was proposed long ago by Cabibbo and Gatto~\cite{eeTL}. 
The $e^{+}e^{-} \to B\bar{B}$ (and its inverse) reaction offers 
new opportunities for studying valence quark effects, diquark pairs (diquarks), 
and the role of different quark compositions~
\cite{eeTh1,eeTh2,eeTh3,eeTh4,eeTh5,eeTh6,eeTh7}. 
This time-like form factor appears to be a viable tool for 
determining hyperon structures, both near the threshold and in the larger 
$q^2$ region, where perturbation effects are expected to be dominant~
\cite{eeex2,eeTL,eeTh6,eeTh7,eeex3,eeex4,eeex5,eeex6}. 
In recent years, 
$e^{+}e^{-} \to B\bar{B}$ and $p^{+}p^{-} \to B\bar{B}$ 
experiments~\cite{expppBB,BESIII3,BESIII4,BESIII5,BESIII6} 
have been able to detect the structure of short-lived 
baryons in the timelike kinematic region above the threshold $4M_{B}^{2}$ 
(where $M_B$ is the baryon mass). 
The experiments at BaBar~\cite{BaBar}, BESIII~\cite{BESIII1,BESIII2} 
and CLEO~\cite{eeTh6,eeTh7} have also provided data related to the electromagnetic 
form factor of timelike hyperons. These timelike studies complement our 
understanding of the spacelike region ($q^2 \leq 0$) \cite{SLTh1,SLTh2,SLTh3} 
based on electron scattering experiments over the past two decades.

The light front quark model (LFQM), a phenomenological model previously 
used extensively to study the form factors of meson weak decays, 
operates not only directly in the time-like regime but also in the space-like regime. 
Using the ``+" component of the current in the 
Drell-Yan-West ($q^{+}=q_{0}+q_{z}=0$) framework, 
the form factors of decays in the space-like regime ($q^{2} < 0$) can be calculated, 
along with contributions from valence and Fock states. Another option 
is to use the same ``+" component but in the $q^{+}\neq 0$ regime, 
i.e., the timelike regime ($q^{2} > 0$). Physically, 
particle decay processes occur in the time-like regime, 
so in principle, the form factors can be calculated in the $q^{+}\neq 0$ 
framework in the time-like regime. However, operating in the time-like 
regime may yield contributions from non-valence Fock states \cite{Brodsky}, 
requiring the wave function to have higher Fock states. 
This means that we will inevitably encounter nonvalence diagrams arising 
from the production of quark-antiquark pairs. 
As shown in Fig. 1(b) and (c), the collision process $e^{+}e^{-} \to B\bar{B}$ 
is a typical non-valence diagram. Fortunately, the literature already 
provides effective methods for dealing with non-valence contributions 
in meson-incompatible processes \cite{Brodsky,CJ1,CJ2,cccq}. The goal of our 
method is to use this procedure to directly calculate the form factor 
in the $e^{+}e^{-} \to B\bar{B}$ collision process and compare it with the experimental data.

This paper is organized as follows. 
In Sec. II, we present the framework for the form factors of $e^{+}e^{-} \to B\bar{B}$.
Our numerical results and discussions are given in  Sec. III. We conclude in Sec. IV.

\section{A framework for $e^{+}e^{-} \to B\bar{B}$}

The reaction where a lepton pair $\l^{+}\l^{-}$ into a $B\bar{B}$ pair, viz
\be
\l^{+}\l^{-}\to\gamma^* \to B\bar{B}\,,
\label{eeBB}
\ee
is a direct source of information on nuclear 
form factors in the time-like region. These is a one-photon-exchange 
for the electron-positron annihilation into a $B\bar{B}$ pair 
and is produced only for $q^{2}\geq 4M^{2}_{B}$.
In particular, to investigate the process of Eq.~(\ref{eeBB}), 
the matrix elements of the nucleon current operator involved 
in this reaction are written as follows,~\cite{crosssec,crosssec2}
\be
\sqrt{\frac{E E'}{M_{B}^{2}}}(2\pi)^{3}
\langle B(P,S) \bar{B}(P',S')|J_{em}^{\mu}(0)|0\rangle=
\bar{u}_{S}(P)
\Big[\gamma^{\mu} f_{1}(q^{2}) 
+i \frac{f_{2}(q^{2})}{2M_{B}}\sigma^{\mu\nu}q_{\nu} \Big]
v_{S^{\prime}}(P^{\prime})\,,
\label{FFs}
\ee
where $q^{\mu}=P^{\mu}+P^{\prime\mu}$ 
and $f_i(i=1,2)$ are the baryonic form factors.
In the time-like region, the current operator participates in 
transitions from vacuum to a state containing hadron pairs, 
which becomes a real state beyond the intrinsic threshold given by 
$q^{2}=4M^{2}_{B}$. 

Employing the single-photon approximation, 
the form shown in Eq.~(\ref{FFs}) is the most general possible form, 
but it is not the only concise expression. 
For the baryon pair production differential cross section, 
after transforming from the center-of-mass frame to the laboratory 
frame, we obtain:~\cite{eeTL,crosssec,crosssec2},
\be
\frac{d \sigma}{d \Omega}\rvert_{(\l^{+}\l^{-}\to B\bar{B})}=
\frac{\alpha^{2}\beta}{4 q^{2}}\bigg\{(1+\cos^{2} \theta)\,|G_{M}(q^{2})|^{2}
+\frac{\sin^{2} \theta}{\tau}\,|G_{E}(q^{2})|^{2}\bigg\}\,,
\label{dcrs}
\ee
where $\alpha$ is fine structure constant, $\beta=\sqrt{1-\frac{4M_{B}^{2}}{q^{2}}}$, 
$\tau= \frac{q^{2}}{4M_{B}^{2}}$, $\theta$ is the angle between 
the direction of the incoming electron and the produced hardron, 
$G_{E}(q^{2})$ and $G_{M}(q^{2})$ are the 
charge and magnetic form factor, respectivly. In comparison to Eq.~(\ref{FFs}), one has 
\be
G_{E}=f_{1}(q^2)+\tau f_{2}(q^2),~~~~~~G_{M}=f_{1}(q^2)+f_{2}(q^2)\,.
\ee

Compared to the traditional form factors $f_{1}(q^{2})$ and $f_{2}(q^{2})$, 
the simplified charge form factor $G_{E}(q^{2})$ and magnetic form factor $G_{M}(q^{2})$ 
provide greater clarity for discussing experimental data.
Integrating the solid angle $\Omega$ over the expression Eq.~(\ref{dcrs}), one obtains
\be
\sigma_{(\l^{+}\l^{-}\to B\bar{B})}=
\frac{4\pi\alpha^{2}\beta}{3 q^{2}}\bigg\{|G_{M}(q^{2})|^{2}
+\frac{|G_{E}(q^{2})|^{2}}{2\tau}\bigg\}\,.
\ee
Therefore, an effective form factor can be defined as follows:
\be
|G_{eff}|=\sqrt{\frac{2\tau |G_{M}(q^{2})|^{2}+|G_{E}(q^{2})|^{2}}{1+2\tau}}\,.
\ee
According to Eq.~(\ref{FFs}), to calculate the form factors, 
we employe a light-front quark model in the time-like region. 
In the LFQM, we treat baryons containing three quarks as bound states 
composed of a single quark $q_1$ and a diquark $q_{[2,3]}$, 
wherein the diquark $q_{[2,3]}$ comprises $q_2$ and $q_3$.
Explicitly, the baryon bound state with the total momentum 
$P$ and spin $S=$$1\over 2$ can be written as~\cite{diquark2}
\be
|B(P,S,S_{z})\rangle & = & \int\{d^{3}p_{1}\}
\{d^{3}p_{[q_2, q_3]}\}2(2\pi)^{3}\delta^{3}(\tilde{P}-\tilde{p}_{1}-\tilde{p}_{[q_2, q_3]})\nonumber \\
&  & \times\sum_{\lambda_{1},\lambda_{2}}\Psi^{SS_{z}}
(p_{1},p_{[q_2, q_3]},\lambda_{1},\lambda_{2})|q_{1}(p_{1},\lambda_{1})[q_2, q_3]
(p_{[q_2, q_3]},\lambda_{2})\rangle\,,
\label{boundstate}
\ee
where $q_{1}$ denotes the active quark of the baryon, $[q_2, q_3]$ represents the diquark, 
$\Psi^{SS_{z}}$ corresponds to the momentum-space wave function and 
$p_{1,2}$  are the on-mass-shell light front momenta,
\be
           p=(p^+, p_\bot)~, \quad p_\bot = (p^1, p^2)~,
                \quad p^- = {m^2+p_\bot^2\over p^+}
\ee with \be
        && p^+_1=x_1 P^+, \quad p^+_{[q_2, q_3]}=x_2 P^+, \quad x_1+x_2=1\,,\nn \\
        && p_{1\bot}=x_1 P_\bot+k_\bot, \quad p_{[q_2, q_3]\bot}=x_2
        P_\bot-k_\bot\,.
\label{Pfraction}
\ee
In Eq.~(\ref{Pfraction}), $(x,k_\perp)$ are the light-front relative momentum
variables, and $\vec{k}_\perp$ is the component of the internal
momentum $\vec{k}=(\vec{k}_\perp,k_z)$.

By the Melosh transformation~\cite{Melosh1974}, 
it is more convenient to work with the following 
representation of the wave function
\be
\Psi^{SS_{z}}(p_{1},p_{[q_2, q_3]},\lambda_{1},\lambda_{2})=
\frac{1}{\sqrt{2(p_{1}\cdot P+m_{1}M_{0})}}\bar{u}(p_{1},\lambda_{1})
\Gamma_{l,m} u(P,S_{z})\phi(x,k_{\perp})\,,
\label{1/2}
\ee
where $\Gamma_{l,m}$ is the coupling vertex function of the decaying quark $q_{1}$
and the diquark in the baryon state.  For the scalar diquark, the coupling vertex is $\Gamma_{s}=1$. 
If the axial-vector diquark is involved, the vertex should be
\be
\Gamma_{A} & =&-\frac{1}{\sqrt{3}}\gamma_{5} \strich\epsilon^{*}(p_{[q_2, q_3]},\lambda_{2})\,.
\label{vertex_1/2}
\ee

The wave function of $\phi(x,k_{\perp})$ in Eq.~(\ref{1/2}) 
describes the momentum distribution of the constituent quarks in
the bound state. In this work, we use the Gaussian-type function, given as
\be
\phi(x,k_{\perp})=4\left(\frac{\pi}{\beta^{2}}\right)^{3/4}\sqrt{\frac{dk_{z}}{dx}}\exp
\left(\frac{-\vec{k}^{2}}{2\beta^{2}}\right)\,,
\label{DA}
\ee
where $\beta$ is the baryon shape parameter and
$k_z$ is defined by
\be
k_z=\frac{xM_0}{2}-\frac{m^2_{2}+k^2_{\perp}}{2xM_0}\,\,,   \quad  M_0=e_1+e_2
\ee
and
\be
M_0^2&=&{ m_{1}^2+k_\bot^2\over 1-x}+{ m_{2}^2+k_\bot^2\over  x}\,,
\ee
where $e_i = \sqrt{m_i^2 + \vec k^2} \,$ and $\vec k =(k_{\bot}, k_z)$. 
In our calculation, we assume
\be
\zeta\equiv{ P^{+}\over q^{+}}
=\frac{q^{2}+M^{2}_{B}-M^{2}_{\bar{B}}-\sqrt{(q^{2}+M^{2}_{B}-M^{2}_{\bar{B}})^{2}
-4 M^{2}_{B} q^{2}} }{2 q^{2}}\,.
\label{eq21}
\ee
Using the bound states of 
$|B(P,S,S_z)\rangle$ and 
$|\bar{B}(P^{\prime},S^{\prime},S_z^{\prime})\rangle$
in Eq.~(\ref{boundstate}) and the above identities,
we derive the matrix elements of the baryonic transition
in the LF frame. By considering the $\mu=+$ component, 
the transition matrix elements are given by
\be
&&\langle B(P,S,S_z)\bar{B}(P^{\prime},S^{\prime},S_z^{\prime})|
\bar q\gamma^{+} q|0\rangle\nonumber \\
&=&
N_{fs}\int{\{d^{4}p_{2}\}}
\frac{\Lambda_{B}(x,{\bf k}_{\perp}) I^{+}
\Lambda^{\prime}_{\bar B}(x',{\bf k'}_{\perp})}{(p_{1}^{2}-m_{1}^{2}+i\epsilon)
(p_{1}^{'2}-m_{1}^{'2}+i\epsilon)}\,,
\label{matrix}
\ee
where $I^{+}=\sum_{\lambda_{2}} \bar{u}(P,S_{z})
\left[\bar{\Gamma}^{\prime}_{S(A)}(\strich p_{1}^{\prime}+m_{1}^{\prime})
\gamma^{+}(\strich p_{1}+m_{1})\Gamma_{S(A)}\right]v (P',S'_{z})$, 
$\Lambda_{B}$($\Lambda^{\prime}_{\bar B})$ corresponds to the vertex function of the 
baryon $B$($\bar{B})$, 
$\bar \Gamma=\gamma^0 \Gamma^\dagger\gamma^0$ and 
$N_{fs}$ is a flavor-spin factor given in Ref.~\cite{Nfs}. 
The trace term $I^{+}$ in Eq.~(\ref{matrix}) can be written as 
the sum of $I^{+}_{1}$ and $I^{+}_{2}$, as shown in Fig. 1. 
\begin{figure}[h]
\includegraphics{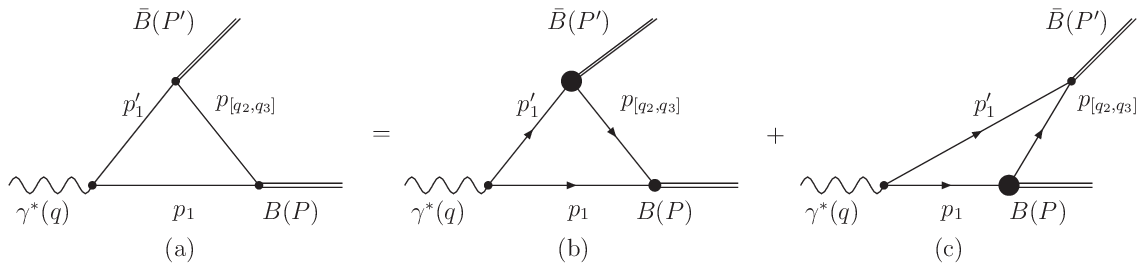} 
\vskip 4cm 
\caption{The effective treatment of the LF amplitude (a) can be displayed into 
two parts in the regions of (b) 
$0 < x < 1$ and (c) 
in $1 < x < 1/\zeta$, where 
the small and big dots of the mediator-quark vertices 
in (b) and (c) represent the LF ordinary 
and nonvalence wavefunction vertices, respectively.}
\end{figure}
In the region  of $0<p_{1}^{+}< P^{+}$ and $p^{-}_{1}=p^{-}_{1on}=
(m_{1}^{2}+k^{2}_{\perp})/p^{+}_{1}$, 
the baryon $\bar{B}$ will possess a nonvalence vertex, as seen in Fig. 1(b). 
The effective contribution of the LF amplitude is given by
\cite{LFcal1,LFcal2,LFcal3,LFcal4}
\be
{\cal M}&=&\frac{N_{fs}}{16\pi^3}
\int^{1}_{0}dx\int d^{2}{\bf k}_{\perp}
\frac{\Gamma_g(x,{\bf k}_{\perp})I^{+}_{1}}
{(1-x)(x'-1)P^{+}P^{\prime+}}\Psi_{B}(x,{\bf k}_{\perp})\nonumber\\
&\times&
\int \frac{dy}{y(1-y)}\int d^2{\bf l}_{\perp}
{\cal K}_{\bar{B}}(x,{\bf k}_{\perp};y,{\bf l}_{\perp})
\Psi^{\prime}_{{\bar B}}(y,{\bf l}_{\perp})\,,\nonumber \\
&&I^{+}_{1}=\bar{u}(P,S_{z})\bar{\Gamma}^{\prime}
(\strich p_{1}^{\prime}+m_{1}^{\prime})\gamma^{+}
(\strich p_{1}+m_{1})\Gamma v(P^{\prime},S_{z}^{\prime})\,.
\label{ValM}
\ee
The wave function corresponding to the nonvalence vertex is usually obtainable  
from the Bethe-Salpeter (BS) amplitude in the BS theory~\cite{FVNV,Brodsky}. 
The LF BS equation  is expressed as
\cite{Brodsky,BSEQ2,BSEQ3}
\be
(M^{2}-M^{2}_{0})\Psi^{\prime}_{\bar B}(x_{i},{k}_{i\perp})
=\int [dy][d^2{\bf l}_{\perp}]
{\cal K}(x_{i},{k}_{i\perp};y_{j},{\bf l}_{j\perp})
\Psi^{\prime}_{\bar B}(y_{j},{\bf l}_{j\perp})\,.
\label{LFBS}
\ee
The nonvalence BS amplitudes can be regarded as solutions of Eq.~(\ref{LFBS}). 
For the baryon, the normal and nonvalence BS amplitudes correspond 
to $x < 1$ and $x > 1$, respectively. 
We note that Eq.~(\ref{LFBS}) essentially takes the same form as the 
LF bound-state equation except the difference in kinematics. 
In Fig. 1(b), the nonvalence BS amplitude 
is the analytic continuation of the valence BS amplitude. 
In the LFQM, the relationship between the BS amplitudes of 
two regions is given in Refs.~\cite{Brodsky,CJ1,CJ2}. 
However, for the integral equation of Eq.~(\ref{LFBS}) 
it is necessary to use the nonperturbative QCD method to obtain the kernel. 
Following Refs.~\cite{LFcal1,LFcal2,LFcal3,LFcal4}, 
we obtain the transition form factors:
\be
f_{1}(q^{2})&=&\frac{N_{fs}}{16\pi^3}
\int^{1}_{0}dx
\int d^{2}{\bf k}_{\perp}\frac{\Gamma_g(x,{\bf k}_{\perp})I^{+}_{f_{1}}}
{(1-x)(x'-1)P^{+}P^{\prime+}}\Psi_{B}(x,{\bf k}_{\perp})\nonumber\\
&\times&
\int \frac{dy}{y(1-y)}\int d^2{\bf l}_{\perp}
{\cal K}_{\bar{B}}(x,{\bf k}_{\perp};y,{\bf l}_{\perp})
\Psi^{\prime}_{{\bar B}}(y,{\bf l}_{\perp})\,,\nonumber \\
I^{+}_{f_{1}}&=&{\rm Tr} [ (\slashed P^{\prime}+M^{\prime})\gamma^{+}(\slashed P+M)
(\slashed p_{1}^{\prime}+m_{1}^{\prime})\gamma^{+}
(\slashed p_{1}+m_{1}) ]\,, \nonumber \\
\frac{f_{2}(q^{2})}{2 M_{B}}&=&\frac{N_{fs}}{16\pi^3}
\int^{1}_{0}dx
\int d^{2}{\bf k}_{\perp}\frac{\Gamma_g(x,{\bf k}_{\perp})I^{+}_{f_{2}}}
{(1-x)(x'-1)P^{+}P^{\prime+}q^{\nu}_{\perp}}\Psi_{B}(x,{\bf k}_{\perp})\nonumber\\
&\times&
\int \frac{dy}{y(1-y)}\int d^2{\bf l}_{\perp}
{\cal K}_{\bar{B}}(x,{\bf k}_{\perp};y,{\bf l}_{\perp})
\Psi^{\prime}_{{\bar B}}(y,{\bf l}_{\perp})\,,\nonumber \\
I^{+}_{f_{2}}&=&{\rm Tr} [ (\slashed P^{\prime}+M^{\prime})\sigma^{\nu +}(\slashed P+M)
(\slashed p_{1}^{\prime}+m_{1}^{\prime})\gamma^{+}
(\slashed p_{1}+m_{1}) ]\,\,,
\label{ffscalar}
\ee
where $\nu = 1, 2$ and the trace term in Eq.~(\ref{ffscalar}) should be separated into 
the on-shell propagating and instantaneous parts of 
$I^{\mu}_{on}$ and $I^{\mu}_{inst}$ via
\be
\slashed p+m=(\slashed p_{on}+m)+\frac{1}{2}\gamma^{+}(p^{-}-p^{-}_{on})\,,
\label{eq24}
\ee
respectively. Both $I^{\mu}_{on}$ and $I^{\mu}_{inst}$ are components of the trace term, 
originating from the first and second terms 
on the right hand side of Eq.~(\ref{eq24}), respectively. 
The explicit forms of the above form factors 
can be expressed as functions of the internal variables of $x$ and $k_{\perp}$. 
In Eq.~(\ref{ffscalar}), $\Gamma_g$ is a vertex function for the gauge boson 
associated with the LF
energy denominator with its explicit form  given by~\cite{Brodsky,CJ2,LFg}
\be
\Gamma_g^{-1}(x,{\bf k}_\perp)=
(1-\zeta)\biggl[q_{\bot}^2 -
\biggl(\frac{{\bf k}^2_\perp + m^2_1}{\zeta (1-x)}
+\frac{{\bf k'}^2_\perp + m'^{2}_{1}}{1 -\zeta-\zeta x}\biggr)
\biggr].
\label{gaugeWF}
\ee
The trace terms in Eqs.~(\ref{ValM}) and (\ref{ffscalar}), 
both corresponding to the products of the initial and final LF spin wave functions, 
can be obtained by off-shell Melosh transformations. 
The form factors related to Fig. 1(b) are given by

\be
f_{1}(q^{2})&=&\frac{N_{fs}}{16\pi^3}
\int^{1}_{0}dx
\int d^{2}{\bf k}_{\perp}\Gamma_g(x,{\bf k}_{\perp})\Psi_B(x,{\bf k}_{\perp}) \nonumber \\
&\times&\frac{[k_{\perp}\cdot k_{\perp}^{\prime}+((1-x_{1})M_{0}+m_{1})
((1-x_{1}^{\prime})M^{\prime}+m_{1}^{\prime})]+I^{+}_{inst}}
{(1-x)(x'-1)}\nonumber\\
&\times&
\int \frac{dy}{y(1-y)}\int d^2{\bf l}_{\perp}
{\cal K}_{\bar{B}}(x,{\bf k}_{\perp};y,{\bf l}_{\perp})
\Psi^{\prime}_{\bar B}(y,{\bf l}_{\perp})\,,
\nonumber
\ee
\be
\frac{f_{2}(q^{2})}{2 M_{B}}&=& \frac{N_{fs}}{16\pi^3}
\int^{1}_{0}dx
\int d^{2}{\bf k}_{\perp}\Gamma_g(x,{\bf k}_{\perp})\Psi_B(x,{\bf k}_{\perp})\nonumber\\
&\times&\frac{
[(m_{1}+(1-x_{1})M_{0})k_{\perp}^{\prime}\cdot q_{\perp}
-(m_{1}^{\prime}+(1-x_{1}^{\prime})M^{\prime})k_{\perp}\cdot q_{\perp}]+I^{\prime +}_{inst}}
{(1-x)(x'-1)q_{\perp}^{2}}\nonumber\\
&\times&
\int \frac{dy}{y(1-y)}\int d^2{\bf l}_{\perp}
{\cal K}_{\bar{B}}(x,{\bf k}_{\perp};y,{\bf l}_{\perp})
\Psi^{\prime}_{\bar B}(y,{\bf l}_{\perp})\,,
\label{f2NV}
\ee
where the instantaneous component $I_{inst}^{(\prime)+}$ of the 
non-valence diagram in Eq.~(\ref{f2NV}) is zero during this process.

The relevant operator ${\cal K}({\cal K}_{\bar{B}})$ 
in Eqs.~(\ref{LFBS}), (\ref{ffscalar}) and (\ref{f2NV}) 
is the BS core, 
which in principle contains contributions from high Fock states. 
It is the high Fock component of the bound state 
related to the lowest Fock component with this kernel. 
The kernel functions can be obtained by using the non-perturbative QCD.
The kernel ${\cal K}$ 
is a function of  all internal momenta ($x,{\bf k}_{\perp},y,{\bf l}_{\perp}$). 
We define that $G_{B \bar{B}}\equiv\int[dy][d^2{\bf l}_{\perp}]
{\cal K}_{\bar{B}}(x,{\bf k}_\perp;y,{\bf l}_\perp)\Psi_{\bar B}(y,{\bf l}_{\perp})$, which
depends only on $x$ and ${\bf k}_{\perp}$. 
The range of the momentum fraction $x$ relies on the 
external momenta for the embedded states. 

In the region of $1 < x < 1/\zeta$, $P^{+}<p_{1}^{+}<q^{+}$ and 
$p^{\prime -}_{1}=p^{\prime -}_{1on}=
(m_{1}^{\prime 2}+k^{\prime 2}_{1\perp})/p^{\prime +}_{1}$. 
At this point, the vertex of the baryon $B$ enters the non-valence region, 
while the vertex of baryon $\bar{B}$ the valence region,  
as shown in Fig. 1(c). For the transition form factors, 
we may directly exchange the following parameters in Eq.~(\ref{ffscalar}): 
\be
x \leftrightarrow x^{\prime},\,~~~~m_{1} \leftrightarrow m^{\prime}_{1},\,
~~~~p_{1} \leftrightarrow p^{\prime}_{1},\,~~~~\Psi_{{B}} 
\leftrightarrow \Psi^{\prime}_{{\bar B}}.
\ee

\se{Numerical Results And Discussions}

We present numerical results on the form factors of the 
$e^{+} e^{-}\to B\bar{B}$ transitions in the time-like region for the LFQM. 
Computationally, we hope to find a time-domain-like result, 
so we only introduce $q_{\perp}= 0$ in the final numerical calculation.
In our calculations, we use ~\cite{diquark2}
\be
m_{u,d} &=& 0.25\,,~ m_{s} = 0.38\,,~\, 
m_{[qq']}=0.7\,,~\, \beta_{s[qq]}= 0.4\pm0.05,
~\beta_{u[qq]}= 0.2\pm0.1\,.
\label{quarkmass}
\ee
In Eq.~(\ref{f2NV}), $G_{B \bar{B}}$ is taken as 
a constant in the range of $0.1 \sim 6.0$, which was previously tested in 
some exclusive semileptonic decay processes and shown 
to be a good approximation for processes with a small momentum transfer~\cite{Brodsky,CJ2}. 
Explicitly,  we choose
$G_{B \bar{B}}=1.8$, 1.0 and 0.3 for 
$\Lambda \Lambda$, $\Sigma^{+} \Sigma^{+}$ and $\Xi^{+} \Xi^{+}$, respectively,
in our numerical evaluation.

To describe the momentum $q^2$ behaviors,
we parametrize the form factors in the double-pole forms of
\be
F(q^2)=\frac{F(0)}{1+a(q^2/m_{P}^{2})+b(q^4/m_{P}^{4})}
\ee
with $m_P=6.0$ GeV for all modes of $e^{+} e^{-}\to B\bar{B}$ 
in $M_{B}$ mass scalar,  
and $[F(0),a,b]$ to be determined in the numerical analysis. 
Our results for the form factors are given in Table I. 
In our results, the errors come from uncertainties in the parameters of $\beta$.
\begin{table}[htbp]
\caption{Form factors of the $e^{+} e^{-} \to B\bar{B}$ transition}
\vskip 0.2in
\label{Table1}
\begin{tabular}{|c||c|c|c|} \hline
$e^{+} e^{-}\to \Lambda \Lambda$ & $F(q^2 = 5.7408\,\,GeV^2 )$ & $a$ & $b$ 
\\ \hline
$G_{E}$& $0.456^{+0.112}_{-0.074}$ 
& $-193.38$ & $1391.17$ 
\\ \hline
$G_{M}$ & $0.462^{+0.113}_{-0.078}$ 
& $-157.969$ & $1133.79$ 
\\ \hline \hline
$e^{+} e^{-}\to \Sigma^{+} \Sigma^{-}$ & $F(q^2 = 6.0\,\,GeV^2 )$ & $a$ & $b$
\\ \hline
$G_{E}$ & $0.0451^{+0.0041}_{-0.0003} $
& $-14.716$ & $91.45$
\\ \hline
$G_{M}$ & $0.0512^{+0.0036}_{-0.0002} $
& $-13.478$ & $83.93$
\\ \hline \hline
$e^{+} e^{-}\to \Xi^{+} \Xi^{-}$ & $F(q^2 = 7.0\,\,GeV^2 )$ & $a$ & $b$
\\ \hline
$G_{E}$ & $0.0903^{+0.0237}_{-0.0156} $
& $-74.80$ & $384.711$
\\ \hline
$G_{M}$ & $0.0965^{+0.0245}_{-0.0164} $
& $-53.141$ & $273.371$
\\ \hline
\end{tabular}
\end{table}

\begin{figure}[t!]
\includegraphics[width=3.1in]{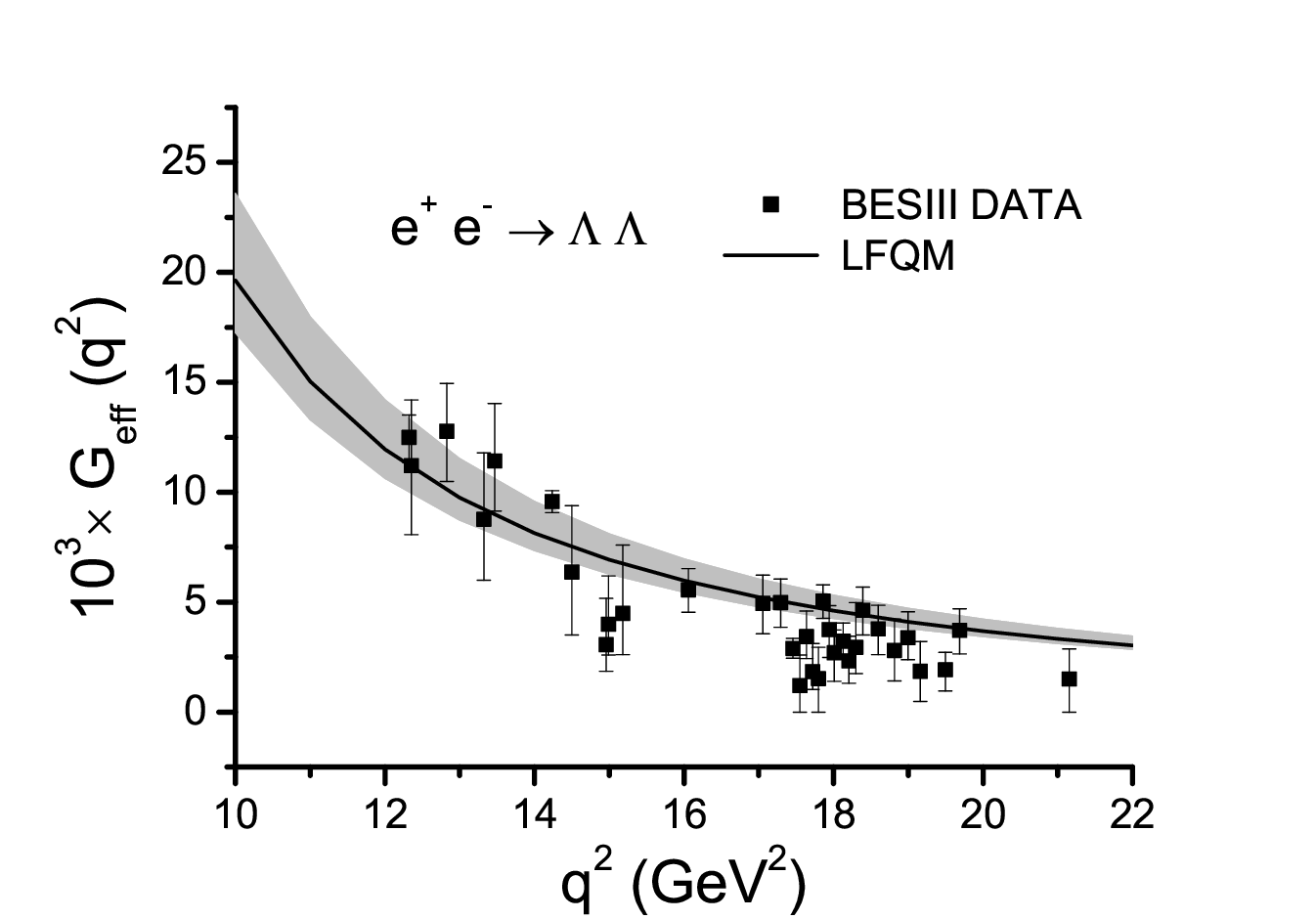}
\includegraphics[width=3.1in]{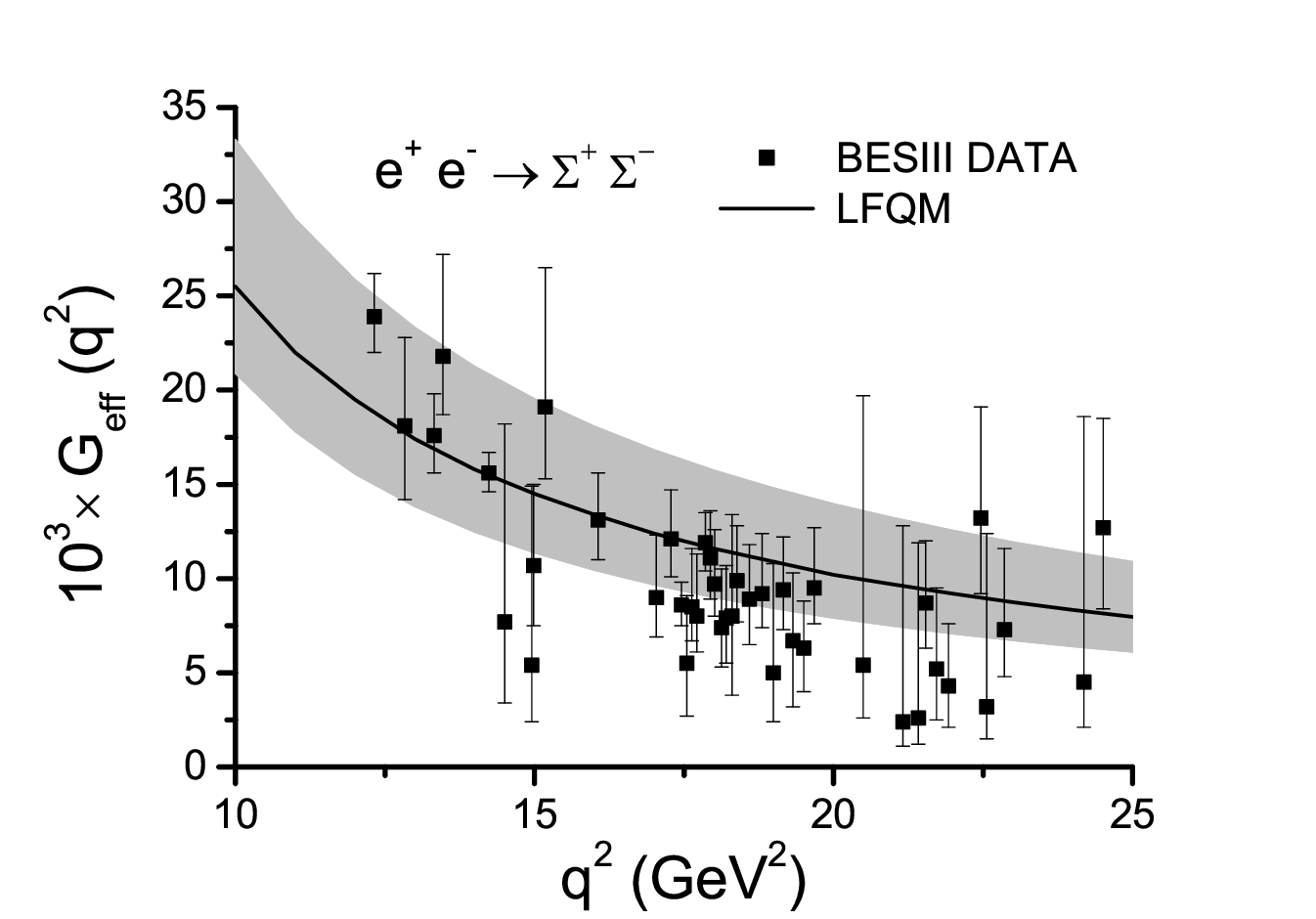}
\includegraphics[width=3.1in]{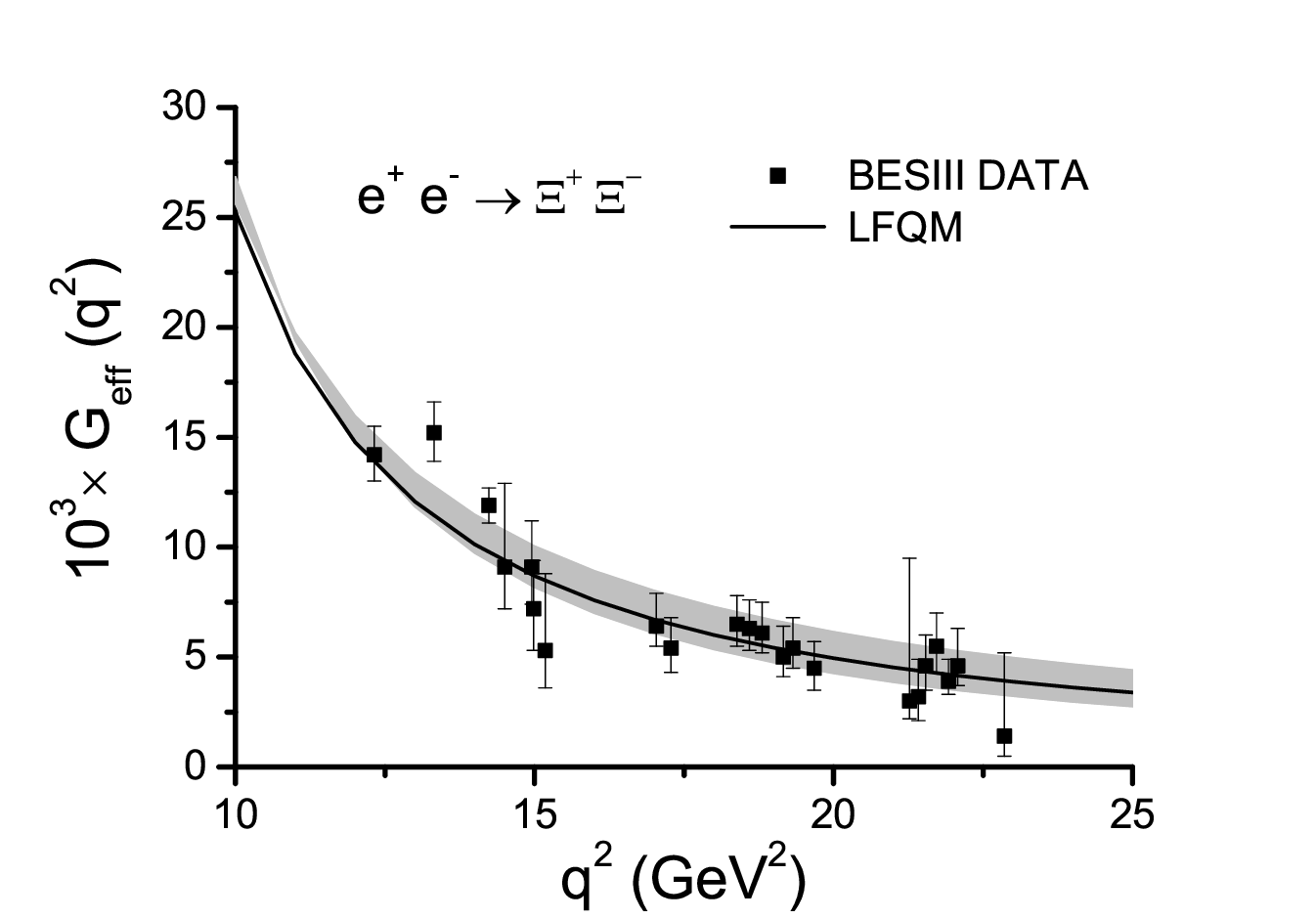}
\caption{Form factors of $e^{+} e^{-}\to B\bar{B}$.}
\label{fig2}
\end{figure}

In Fig.~\ref{fig2}, we present our evaluations of 
the form factors as functions of $q^2$ for the
$e^{+} e^{-}\to \Lambda \Lambda, \Sigma^{+} \Sigma^{-}$ 
and $\Xi^{+} \Xi^{-}$ transitions, 
respectively, where the BESIII data fits are also given. 
From the figures, we compare the LF model 
calculations for $|G_{eff}|$ with the BESIII data 
of $\Lambda\Lambda, \Sigma^{+} \Sigma^{-}$, 
and $\Xi^{+} \Xi^{-}$ for $q^{2} \geq 10$ GeV$^2$. 
This comparison shows that 
the LF model describes the data well above 
$q^2 \geq 10$ GeV$^2$ within the theoretical limit. 
Current calculations for $|G_{eff}|$ indicate that the region $q^2 \geq 10$ GeV$^2$ 
is within the range, where the asymptotic behavior of the form factor can be observed. 
However, it is important to note that the current data may still be in 
the non-perturbative QCD regime.

Fig.~\ref{fig2} shows that the form factors exhibit a similar 
behaviors throughout the $q^{2}$ spectra.  
Obviously, our LFQM results with the non-valence contributions 
are consistent with the BESIII data~\cite{BESIIIL,BESIIIS,BESIIIX}. 
We also use this model to calculate the $|G_E/G_M|$ ratios 
for the processes 
$e^{+} e^{-}\to (\Lambda \bar{\Lambda}$,\,$\Sigma^{+} \Sigma^{-},\,\,\Xi^{+} \Xi^{-}$).
For the three processes mentioned above, at $q^2 = (5.7408,6.0,7.0)$ GeV$^2$, 
we get 
$|G_{eff}| = (0.921^{+0.22}_{-0.15},~0.098^{+0.002}_{-0.0002},~0.189^{+0.048}_{-0.032})$ and 
$R =|$$G_E\over G_M$$|= (0.985\pm 0.001,~0.88\pm 0.01,~0.936\pm 0.004)$.

\se{Conclusion}

The $e^{+}\,e^{-}$ collision experiments explore the electromagnetic properties 
of elementary particles. Under the electromagnetic interaction at the tree level, 
$e^{+}\,e^{-}$ annihilation processes can occur via virtual 
photons with timelike four-momenta, which subsequently decay into final states.
In this work, we have used the LFQM to investigate the 
form factors in the $e^{+}e^{-} \to B\bar{B}$ collision process. 
In particular, we have analyzed the form factors for the baryon 
transition based on the BS formalism with $q^{+} > 0$, 
effectively accounting for nonvalence contributions. 
We have found that the $q^2$ behaviors of the form factors are consistent 
with the BESIII data.  As shown in Table 2, for the processes 
$e^{+} e^{-}\to (\Lambda \bar{\Lambda}$,~$\Sigma^{+} \Sigma^{-},~\Xi^ {+} \Xi^{+})$  
at $q^2$=$(5.74,~6.0,~7.0)$ GeV$^2$, 
our results are $|G_{eff}| = (0.921, 0.098, 0.189)$ 
and $R = |G_E/G_M| = (0.97, 0.89, 0.936)$, 
the results including nonvalence contributions agree with 
existing BESIII experimental data and other theoretical estimates.

\section*{Acknowledgments} 
This work is supported in part by 
the National Natural Science Foundation of China (NSFC) under Grant No. 12547104.

\end{document}